• Article •

# Visualizing Plasma Physics Simulations in Immersive Environments

Nuno Verdelho Trindade[1*], Óscar Amaro[2], David Brás[3], Daniel Gonçalves[1], João Madeiras Pereira[1], Alfredo Ferreira[1]

1. *INESC-ID, Instituto Superior Técnico, University of Lisbon, Rua Alves Redol, 9, 1000-029 Lisbon, Portugal*
2. *Group of Lasers and Plasma (GoLP), Instituto Superior Técnico, University of Lisbon, Complexo Interdisciplinar, Av. Rovisco Pais, 1049-001 Lisbon, Portugal*
3. *Department of Computer Science and Engineering (DEI), Instituto Superior Técnico, University of Lisbon, Av. Rovisco Pais, 1, 1049-001 Lisbon, Portugal*

**Supported by** this work was supported by national funds through Fundação para a Ciência e a Tecnologia (FCT), under grants 2021.07266.BD and UI/BD/153735/2022.

**Abstract Background** Plasma physics simulations create complex datasets for which researchers need state-of-the-art visualization tools to gain insights. These datasets are 3D in nature but are commonly depicted and analyzed using 2D idioms displayed on 2D screens. These offer limited understandability in a domain where spatial awareness is key. Virtual reality (VR) can be used as an alternative to conventional means for analyzing such datasets. VR has been known to improve depth and spatial relationship perception, which are fundamental for obtaining insights into 3D plasma morphology. Likewise, VR can potentially increase user engagement by offering more immersive and enjoyable experiences. **Methods** This study presents *PlasmaVR*, a proof-of-concept VR tool for visualizing datasets resulting from plasma physics simulations. It enables immersive multidimensional data visualization of particles, scalar, and vector fields and uses a more natural interface. The study includes user evaluation with domain experts where *PlasmaVR* was employed to assess the possible benefits of immersive environments in plasma physics visualization. The experimental group comprised five plasma physics researchers who were asked to perform tasks designed to represent their typical analysis workflow. To assess the suitability of the prototype for the different types of tasks, a set of objective metrics, such as completion time and number of errors, were measured. The prototype's usability was also evaluated using a standard System Usability Survey questionnaire. **Results** The findings indicate that the prototype is better fitted for preliminary visualization of the plasma physics data and to handle analytics tasks rather than annotation tasks. Participants manifested a high level of engagement when using the prototype, considering it more enjoyable than conventional means. They found the system usable, with a reliable, well-structured, and cohesive interface. They also considered that the interface was easily learnable. Furthermore, the participant's perception of the usefulness of VR in plasma simulations increased after experiencing the prototype. **Conclusions** Interactive, immersive environments can be applied in a useful and usable manner in plasma physics simulations data analysis. They also have the potential to provide a more enjoyable and engaging experience than conventional visualization means.

**Keywords** Virtual reality; Immersive visualization; Scientific visualization; Plasma physics; Data analysis

# 1 Introduction

Plasma is a physical state of matter where a significant fraction of particles is charged [1]. These particles, usually electrons and ions, interact via long-range forces and sustain rich, collective motion, waves, and instabilities [2]. Plasmas account for the universe's most ordinary, visible matter, including planetary magnetospheres, stars, interstellar space, and accretion disks around neutron stars and black holes [3]. In some scenarios, the highly nonlinear dynamics of this medium hampers a purely theoretical analysis. Modeling these systems often requires calculating the complete six-dimensional phase-space of particles in time, in addition to the self-consistent electromagnetic fields and currents generated by them [4]. These complex simulations can be performed using *Particle-in-Cell* (PIC) codes [5]. An example of such a code is the fully relativistic, massively parallel *OSIRIS PIC* code [6] used to generate the data presented in this work.

Given that plasma physics simulations can result in large datasets with millions of data points and complex structures, they are not trivially understood merely through standard rendering of the numerical results [7]. Therefore, researchers need state-of-the-art visualization tools to gain intuition about the simulated physical phenomenon. The datasets obtained from these simulations consist most notably of particle data (e.g., position, momentum, energy), scalar field data (e.g., energy density), and vector field data (e.g., electric and magnetic fields) [8]. Depending on the exact physical phenomena being visualized, the datasets might have different characteristics: they can be more chaotic (e.g., when simulating plasma turbulence) or have more discernible patterns (e.g., advanced plasma-based particle acceleration concepts) [9]. The generated datasets are 3D in nature but commonly depicted and analyzed by resorting to 2D idioms. While there are applications to create and analyze 3D idioms for plasma physics[10,11], these use conventional visualization and interaction means (2D screens, keyboard, and mouse), which are not particularly engaging[12]. In that sense, virtual reality (VR) can be used as an alternative to these conventional means in the analysis of plasma simulations. VR has been known to offer improved depth and spatial relationship perception [13,14], which are fundamental for obtaining insights into 3D plasma morphology. It provides a different perspective that results from the users' immersion in the physical constructs they are trying to observe [15]. Likewise, VR can potentially increase user engagement by offering more immersive and enjoyable experiences.

This study presents *PlasmaVR,* a VR interactive prototype tool to visualize the data resulting from plasma physics simulations. The tool provides researchers an immersive environment for exploring scientific datasets using natural interaction. It enables multidimensional data visualization of particles, scalar, and vector fields. Users can travel inside animated 3D idioms based on time-dependent data and see the structural variations from several points of view. They can move around distraction-free and work within an extensive virtual data display real estate. The tool includes specific functionalities for data annotation and segmentation. In that sense, it allows multiaxial dynamic slicing of the 3D idioms with a real-time representation of corresponding scalar 2D energy heatmaps.

The design specifications of *PlasmaVR* were based on a series of interviews with domain experts. These experts are researchers of the *Group of Lasers and Plasma* (GoLP) at Instituto Superior Técnico (IST), University of Lisbon. They routinely create plasma simulations using PIC code methods. The datasets used for evaluating the prototype result from real simulations carried out by GoLP. The prototype was also designed to integrate these datasets into immersive environments seamlessly. Such integration required the automation of raw data pre-processing, the creation of idioms based on the resulting parsed information, and the implementation of natural interaction support in the immersive environment.

In addition to presenting *PlasmaVR,* we also intended, with this study, to find out if there were any tangible advantages in using immersive environments for plasma physics visualization. Particularly, we wanted to understand if VR was usable in that scope and could lead to useful, engaging experiences. As such, this study uses the prototype to assess the following research questions:

**RQ1**: Can VR, applied in the context of plasma physics visualization, result in a usable experience?

**RQ2**: Can VR be useful in plasma physics visualization?

In that context, *PlasmaVR* was evaluated with a group of domain experts. They performed a set of relevant tasks with the prototype, using real datasets, which simulated their typical analysis workflow. A collection of relevant objective metrics was registered. The participants were also asked to answer a standard usability questionnaire and provide additional feedback regarding their user experience. Therefore, the major contributions of this study are a **novel tool for plasma physics scientific immersive visualization** and a **discussion on the usefulness of immersive interactive environments in plasma physics visualization**.

## 2 Related work

The use of immersive technologies has been widely accepted as an aspect of paramount relevance and a future trend in scientific visualization [16]. In that scope, extended reality (XR) has been applied in a wide range of domains, including in research studies related to natural, formal, and social sciences [17–20]. Some of this existing work, focused on using XR technologies in the scope of general physics, vector fields, and plasma physics scientific visualization, is exemplified below.

### 2.1 Physics visualization

VR and other XR technologies have been broadly applied in physics visualization. An example is the work developed by Orlando et.al. [21]. They created the *3DMAP-VR* (3-Dimensional Modeling of Astrophysical Phenomena in Virtual Reality) project, which focused on setting up a workflow for VR visualizations of astrophysical phenomena. The workflow aims to make these visualizations easily accessible for scientific analysis and public outreach. It consists of using the data from simulations obtained using parallel codes (similar to the PIC methods used to obtain plasma simulations) and processing this data into 3D visualizations for VR environments. This processing is made using *ParaView* [10] and *MeshLab* [22]. Depending on their significance for the simulated phenomena, these tools are used to create 3D models that combine isosurfaces and streamlines. The 3D models are optimized, appropriately colored, and tuned using these tools and end up occupying less storage space than the datasets being represented. The models can then be visualized using a VR headset. The system is especially effective for public outreach events. In the scope of proton physics, Polys et.al. [23] developed a processing pipeline for producing interactive scientific visualizations of high-dimensional particles. They tested the created visualizations using both a VR headset and a CAVE system. The system also uses *ParaView* as a processing and visualization platform.

## 2.2 Vector fields visualization

Various research work has also tackled the intricacies of vector field visualization and interaction in VR. The study by Kageyama et.al. [24] focused on visualizing vector fields using a *CAVE* system. They implemented streamline and arrow visualizations. The streamline visualizations are created using an initial placement for a particle (the seed placement) and then integrating the particle position using the vector field data, making it possible to obtain the particle movement. The user can interactively create streamlines in the virtual environment using an interface to place seeds. The arrow visualizations are made by placing arrows in certain positions of the vector field, pointing in the specified direction, with length according to the magnitude of the field. Then, the user can move the controller and watch how the field evolves around their hand. Also, regarding vector field visualization, Yoshizaki and Kageyama[25] tackled the problem of visualizing a complex vector field in a way that is not too confusing for the user. They proposed the *COMB* method for placing streamline seeds using a VR controller. The aim is to cast a ray from the controller and then place the seeds at specific intervals within the ray. The streamlines then expand a certain distance in each direction. The number of controlled seeds is small enough to execute the integration at run time, allowing the user to analyze the simulation with the VR controller dynamically.

## 2.3 Plasma physics visualization

The application of XR technologies specifically to plasma physics data visualization has equally been addressed in previous research work. An example is the system developed by Foss et.al. [26], where plasma simulation data was visualized in an augmented reality (AR) environment. This work focuses on the scalar data obtained from plasma simulations and uses isosurfaces for its representation. *Paraview* was, once again, used to process the data, create the visualization, and export it as a *Unity-compatible* file format. The visualization is presented in a time-varying format, and as such, 3D models corresponding to the timesteps were pre-rendered and then displayed in sequence to create an animation. The visualization can be done in an AR environment using, e.g., a *HoloLens*[1] headset. The system is limited in what concerns interactivity, as possible user interaction mainly consists of walking around the visualization. However, researchers who tested the system found the possibility of observing the isosurfaces from different perspectives in an AR environment very valuable. The integration of plasma physics simulations with VR was equally explored by Danielová et.al. [27]. They proposed a system that allows researchers to visualize plasma simulation datasets. The web-based system offers a new perspective on complex interactions of intense laser beams with various forms of targets. It enables the analysis of particles and fields and the modification of environmental properties to enhance spatial features.

Ohtani et.al.,[28–30] used a *CAVE* system for visualizing plasma simulation results and device data. They addressed the virtual representation and interaction with magnetic field lines, particle trajectories, and isosurfaces of plasma pressure. The study focused on the virtues of VR in understanding the three-dimensional positional relationship between plasma elements. In that scope, researchers also addressed in a later work [31] how VR promoted users' understanding of dust particle positioning in plasma simulation experiments. They concluded that VR could improve understanding of the relative positional relationship between the dust particles' trajectories and the magnetic field's structure. In an earlier study, Hayashi [32] had also already conceptualized how *CAVE* systems could be used to simulate nonlinear phenomena in plasmas. Ohno et.al. [33] had previously presented *VFIVE*, a VR visualization software for *CAVE* systems, capable of representing scalar and vector data. Lastly, the application developed by Brás[34], which was used as a base for the current study, uses VR headsets in the scope of scientific visualization of plasma physics simulations.

The present study builds upon visualization and interaction concepts addressed in earlier work and further explores the potential of immersive environments in plasma physics scientific visualization. In particular, it fills in the gaps concerning the usefulness and usability of VR in that scope. While Foss et.al. [26] and Ohtani et.al. [28–30] used CAVE and AR systems, we set out to find out how a system capable of a potentially more immersive experience (designed for VR headsets) could perform in that sense. Also, unlike the study conducted by Foss et.al. [26], we used an interface with integrated analytics features. These features share some of the characteristics of the work developed by Danielová et.al. [27] but were designed natively for VR interaction instead of the web interface proposed by these researchers.

## 3 System overview

Our proposed system, *PlasmaVR,* is an interactive tool for scientific visualization and exploration of datasets that result from plasma physics simulations in an immersive environment. The tool allows users to load raw plasma experiment datasets and visualize and interact with their particles, scalar, and vector fields. It includes specialized analytics functionalities, like field segmentation. It also incorporates 3D data annotation. The system is composed of two complementary modules. The first is a desktop data parsing application that transforms the original raw data from plasma simulation experiments into a format compatible with the graphical engine. The second is the VR application itself, which enables the user to visualize and interact with the 3D idioms in an immersive environment.

---

[1] Hololens headset, https://www.microsoft.com/en-us/hololens

The general requirements for *PlasmaVR* were established from interviews with domain experts from GoLP. For the processing module, the general requirements can be summarized by the following points:

- Ability to process raw particle files. This sequence of files contains data such as position, energy, or momentum at each specific time step. A typical maximum of two million particles for each time step had to be supported;
- Ability to process scalar field files. This sequence of files contains scalar data concerning the simulation in a regular three-dimensional grid at each time step of the simulation. These grids range from 100³ to 500³ in size;
- Ability to process vector field files. This sequence of files contains vector data in a regular three-dimensional grid for each simulation time step. These grids range from 100³ to 500³ in size.

For the VR module, the following general requirements were established:

- Animated visualizations to show the changing dynamics of the simulated system;
- Separate visualizations for each of the different data structures: particle positions, scalar fields, and vector fields;
- A playback control mechanism for the animated visualization;
- Panels to control the visualization parameters and the characteristics of the immersive environment;
- Ability to slice and segment the 3D idioms into 2D energy heatmaps;
- Dynamic time-dependent annotation capabilities.

## 3.1 Architecture

*PlasmaVR* was developed using the *Unity* game engine and the *C#* programming language. These choices allow broad compatibility with distinct VR headset models and fast application development. As previously mentioned, the system aggregates two complementary modules: the data processing module and the VR module. The first one aims to process the high volume of raw data from plasma simulation experiments. This raw data is generated in the *HDF5* [35] format, and the processing module transforms it into the *OBJ*[2] format, which is then used inside the graphical engine to generate the corresponding 3D idioms. This module can process point clouds as well as scalar and vector volumes. The second module handles the representation of the different virtual elements inside the immersive environment and the user interaction. In that scope, it is responsible for rendering the particle clouds, isosurfaces, and streamlines. The prototype's architecture is depicted in Figure 1. In the following sections, the inner workings of these two main modules of *PlasmaVR* will be discussed in greater detail.

---

[2] Specifications of the Wavefront OBJ format: https://www.loc.gov/preservation/digital/formats/fdd/fdd000507.shtml

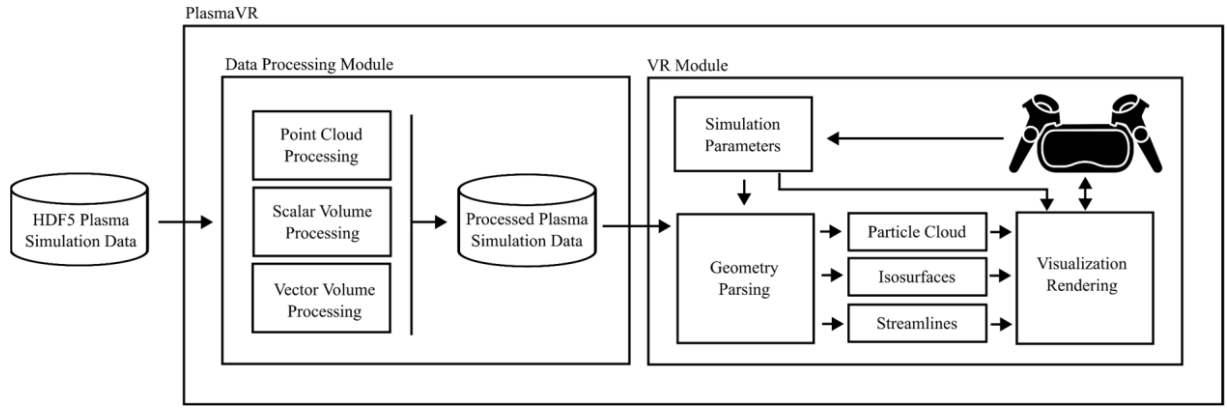

**Figure 1** *PlasmaVR* system architecture.

## 3.2 Data processing and representation

The raw data files from plasma physics experiments cannot be used directly in the graphical engine to represent idioms. This raw data in *HDF5* format must be first processed and outputted to the *OBJ* format. Furthermore, this original data must be abstracted into relevant derived measures to be useful. A desktop application with a GUI, developed in *Python*, was designed explicitly for that purpose. This application can process the different types of raw datasets detailed below.

The particle datasets consist of the raw particle data that results from the plasma simulation. They include each particle's positions and energy values, which consist of four floating-point values. Despite being reasonably simple datasets, they typically have a large size. When creating a time-varying simulation visualization, researchers are trying to visualize the general state of the whole system and how it evolves. Therefore, there are no advantages in displaying the complete set of particles in the dataset. The system's state can be depicted with sufficient accuracy by only showing a representative subset of the particles. Randomly sampling a set proportion of the particles can reduce the visual overlap and allow for the state dynamics of the system to be displayed with lower graphical processing costs.

The scalar field datasets consist of the set of charge density values at each simulation time step. These charge densities can be represented using isosurfaces. An isosurface is a geometric representation of a value in a volume. The surface is generated inside a volume so that all its points have the same value, creating a simplified representation of the value distribution. The processing module addresses isosurface creation by implementing the *Flying Edges* algorithm [36]. This algorithm uses multiple passes to pre-allocate memory for points and primitives, which is then used to achieve faster execution speed.

In a time-varying simulation, a plasma physics researcher will give greater importance to the overall shape of the plasma structure and not the contour details. Therefore, the obtained surfaces are simplified using a mesh decimation algorithm, in this case, the *Quadric Error Mesh Decimation* algorithm [37]. After the simplification, the resulting geometry can finally be exported in *OBJ* format. This file format contains geometric meshes, including UV mapping and normals. The complete scalar field processing sequence is systematized in Figure 2.

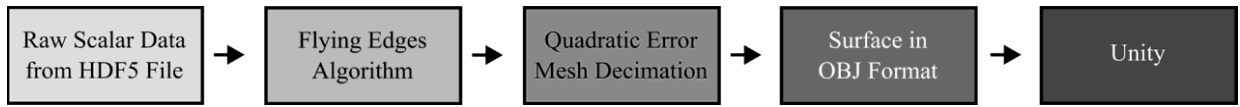

**Figure 2** Steps of scalar field processing.

Similarly to the scalar fields, the vector field datasets are regular three-dimensional grids. However, these fields contain vector information concerning the direction and intensity of the magnetic field at each point in the grid. The vector fields can often be chaotic due to the nature of plasma simulations. Thus, their visualization is complex. A vector field can be represented through various means, such as using arrows at specific points in the grid to indicate the direction of the field. However, given the density of the fields at hand, a streamlined representation was chosen as the preferred method. The streamlines are tangent to the field direction and colored according to the field intensity at that point in the line.

The method chosen to compute these streamlines was the *Runge-Kutta* method [38], a discrete integration method that can be used to obtain the successive points that create the streamline (Figure 3). These points are obtained by applying the field direction at each point, in specific proportions described by the relevant *Runge-Kutta* method, to minimize the error from making such an integration. The position of each point in the line and the magnitude of the vector field at that specific point are then ready to be transported and represented in the immersive environment.

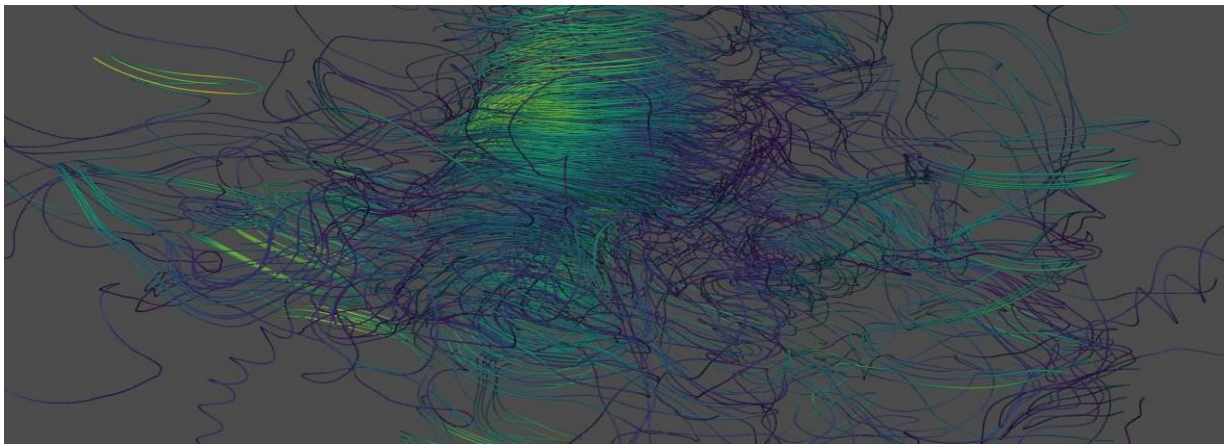

**Figure 3** Representation of the streamlines resulting from the processing of raw data.

Regarding data representation in the graphical engine, the *Unity Visual Effects* (VFX) *Graph*[3] was used to visualize particles. This feature allows the creation of high-quality particle effects while taking advantage of the parallel processing offered by the GPU. Due to the very high number of particles that need to be represented, this optimized graphical processing is invaluable to achieve a good performance in VR.

The particle files are read and converted to textures in *Unity*. Each point of the texture represents a particle, encoding four floating-point values (the three spatial coordinates and the numerical information about the particle that was imported). One of these textures is kept for each of the timesteps, with adjustable resolution (which limits the number of particles that can be displayed). A resolution of 512×512 offered the best results, allowing for the most significant particles to be displayed while maintaining a suitable performance. These textures are used as maps by the VFX Graph to position and color each particle accordingly. The different textures are then swapped in sequence to display the animated particles.

---

[3] Unity VFX Graph: https://unity.com/visual-effect-graph

The scalar and vector fields have similar approaches in their representation but with some differences when importing the data to the VR application. The scalar fields are visualized with the previously generated isosurfaces. The *OBJ* models are imported sequentially and are stored in memory to later create the animation by swapping the displayed model. A custom shader is then applied to the isosurfaces to render both sides of the model. These surfaces act as a boundary for a specific value within the scalar field, and as such, they need to be viewable from outside and from within the plasma structure.

The vector fields are imported differently by using the created streamline data. The line is described with four floating-point values in a sequence. The position of each point of the line, as well as the intensity of the vector field at each point. Then, the streamlines are displayed using a *Unity* mesh object with a line topology. As the line visualization is stored in a mesh object, the method used to display the isosurfaces can also be generalized to the streamlines. Figure 4 represents the idioms shown in the prototype for these three modalities (particles, scalar, and vector fields).

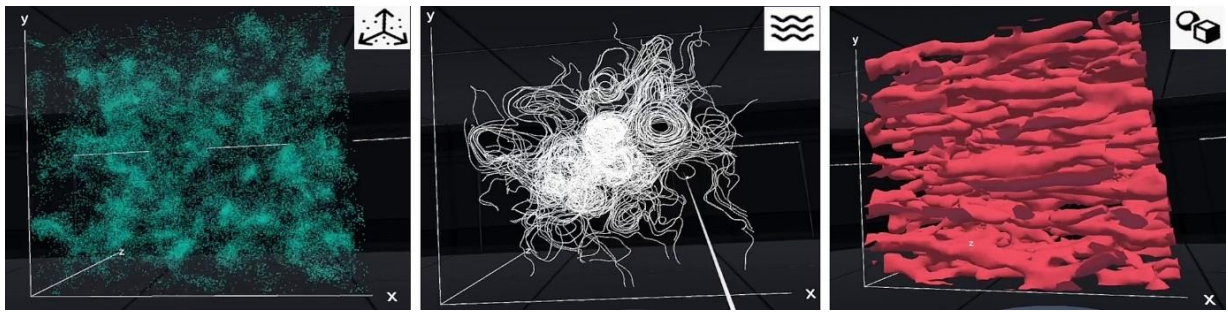

**Figure 4 Particles (left), streamlines (center), and isosurfaces (right) idioms represented in the immersive environment and corresponding symbols used in the interface (top-right corners).**

## 3.3 User interface

When the users put on their VR headsets, they are inserted into an environment that simulates a sci-fi laboratory-themed room. This virtual room is where the plasma exploration takes place (Figure 5). In the center, the chosen type of plasma structure fluctuates above a platform and can be animated, rotated, resized, sectioned, or annotated. The user can move around the room, enter inside the plasma structure (Figure 6), slice it, and observe the intricacies of its morphology.

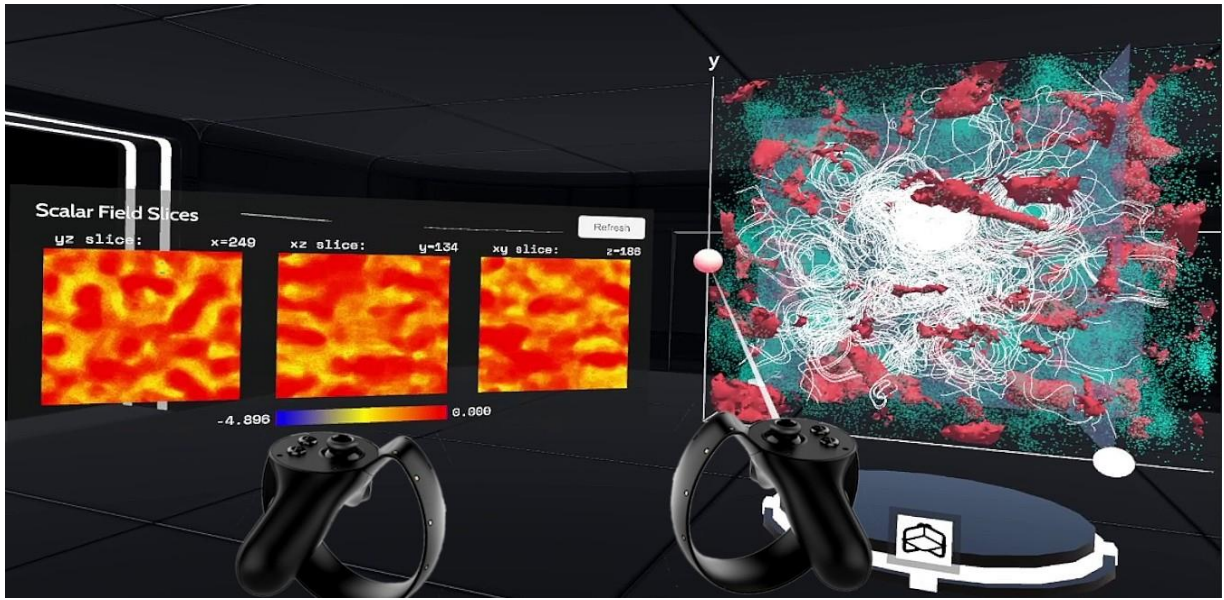

**Figure 5** The immersive environment (mockup) of *PlasmaVR* emulates a sci-fi-themed laboratory. The idiom and the slicing planes control mechanism are visible on the right of the figure. The resulting slices are on the left of the figure.

The continuous movement of the user was chosen as the modality of locomotion in the environment. This type of locomotion allows a more natural exploration of the 3D idioms. The user can move around the room both by physically walking and by using the direction joysticks of the VR controller. Alternatives such as teleportation were passed over due to the relatively small dimensions of the virtual room, or Walking-in-place (WIP) due to the need of an extra depth camera [43].

For accessing the different features, *PlasmaVR* uses a floating panel (which can be shown or hidden) attached to the left VR controller. It includes buttons that lead to the playback control, slicing, and rotation/resizing features, annotation, and to the immersive environment configuration. The first of these features is the playback control, which incorporates play/pause buttons and a timeline. This timeline can be used to jump to a specific frame in the idiom's animation or observe step-by-step modifications of the plasma structure (Figure 6).

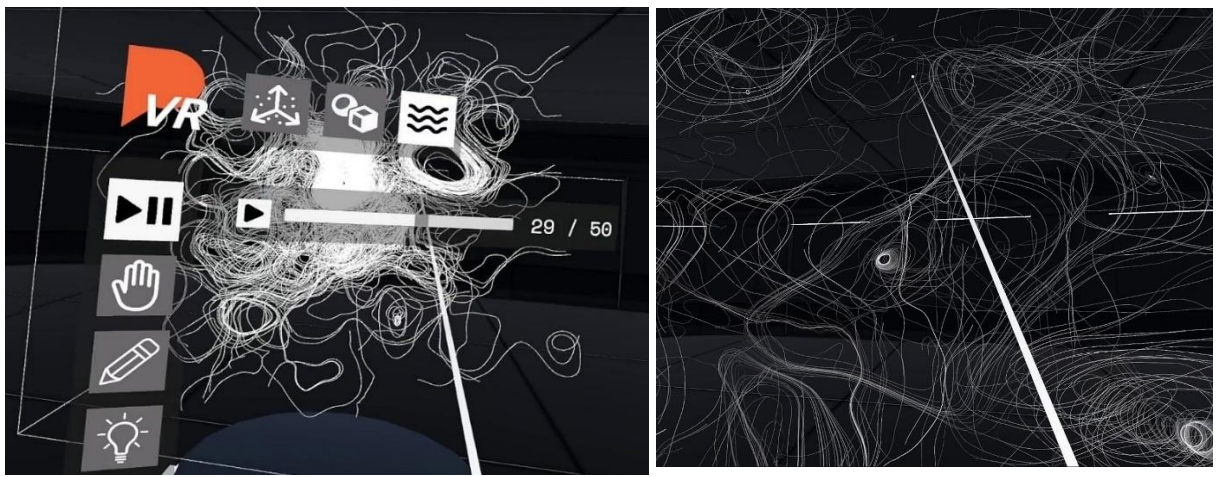

**Figure 6** *PlasmaVR* playback menu (left). The plasma animation can be more precisely controlled by dragging the timeline marker. Entering inside the plasma structure for a better understanding of its morphology (right)

When analyzing the idioms, the users can rotate and resize them to view the data from different angles and perspectives. The rotation of the idioms is performed with a mapping to the controllers' motion. When the rotation mode is activated, a virtual model of a hand holding a sphere appears in place of the model of a standard controller. This sphere works as a 'proxy' for the 3D idiom. The user can then press the controller trigger to grab the sphere and rotate it with their virtual hand (Figure 7). The idiom follows the sphere's rotation with three degrees of freedom.

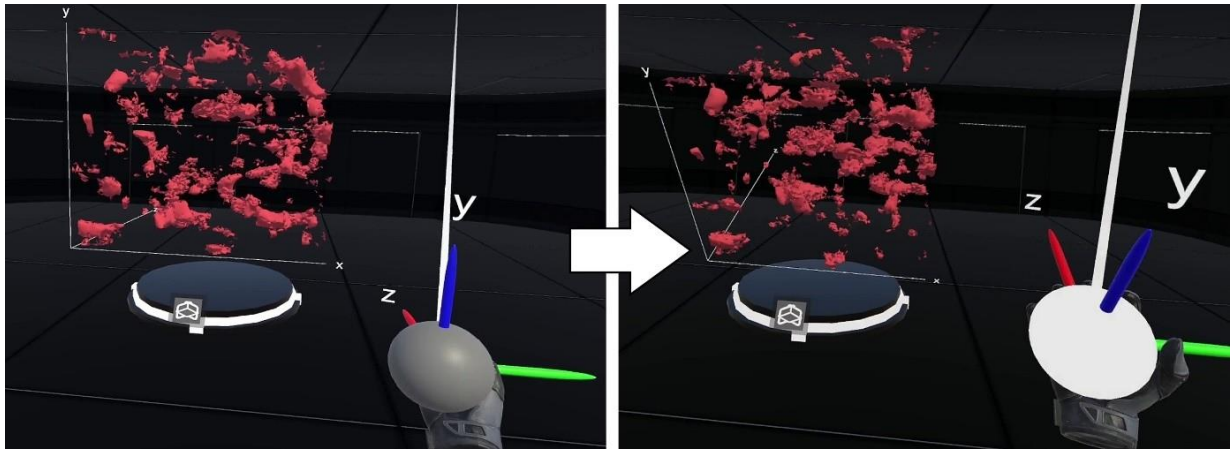

**Figure 7 Controlling the rotation of the 3D idiom using the 'proxy' sphere paradigm. In the figure, the idiom is rotated around the XX axis by imparting the same rotation to the sphere.**

As previously mentioned, following the initial assessment of the application, GoLP researchers suggested that it would be relevant to slice the scalar field using planes to display the energy distribution within that slice. As such, *PlasmaVR* includes a slicing feature, which is directed at producing formal results of data analysis. To extract the slices, the users can move small spherical handles attached to each axis of the 3D plasma idiom. These handles are, in turn, linked to planes that bisect the idiom and will move along its respective axis. The position of each plane determines the sections that will be extracted (Figure 8).

A 2D panel displays the heatmaps corresponding to the bisection from the three planes. To represent the heatmaps, the application uses a large amount of volumetric data, and to improve performance, the current frame of the simulation is loaded to memory. When the new data is loaded, a color gradient is created using the maximum and minimum values available as the edges. This volumetric data is then used to create the textures to display. By evaluating each value in the slice according to the gradient, a colored texture is obtained, mapping the scalar field values in that specific plane. Due to the availability of the data in memory, the slices can be moved in real time, and the researcher can watch the heatmaps change accordingly. This real-time multiaxial representation of energy heatmaps allows the researchers to examine how the values change inside the represented fields.

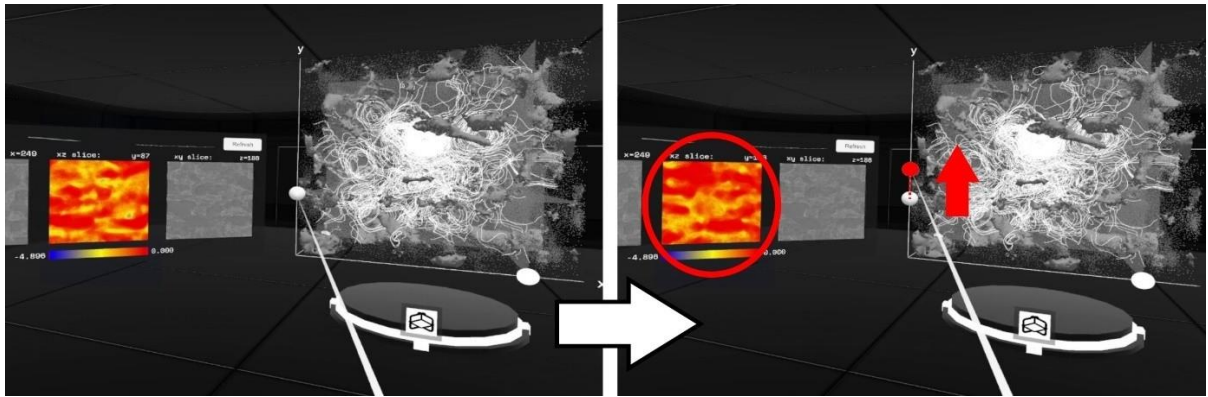

**Figure 8** The three intersecting planes can be moved along its axis to generate the corresponding slices. In the example of the figure, moving the handle of the horizontal plane up the YY axis (red arrow and small red circle) results in the real-time update of the corresponding energy heatmap for the XZ slice (large red circle).

When analyzing a plasma idiom, researchers may want to highlight specific aspects of what is happening in the simulation. However, annotating something in a 3D idiom using a keyboard, mouse, and 2D screen may be tedious and counterintuitive. Thus, researchers usually either make 2D annotations or generate and move 3D objects (e.g., arrows) to the area they want to highlight. On the contrary, 3D annotation fits particularly well within the *PlasmaVR* environment and interaction.

The annotation feature takes advantage of VR's increased depth and spatial relationship perception. The researcher can enter inside the plasma structure and make geometrically complex and precise 3D annotations, thus highlighting valuable simulation insights. Due to the ability to draw freely in 3D, it is easy to draw annotations like in Figure 9, where the purple annotation wraps around a volumetric protuberance in the structure, thus providing information more comprehensively. The annotations can be aggregated in groups, and their colors can be customized using the annotations menu, accessible from the main panel.

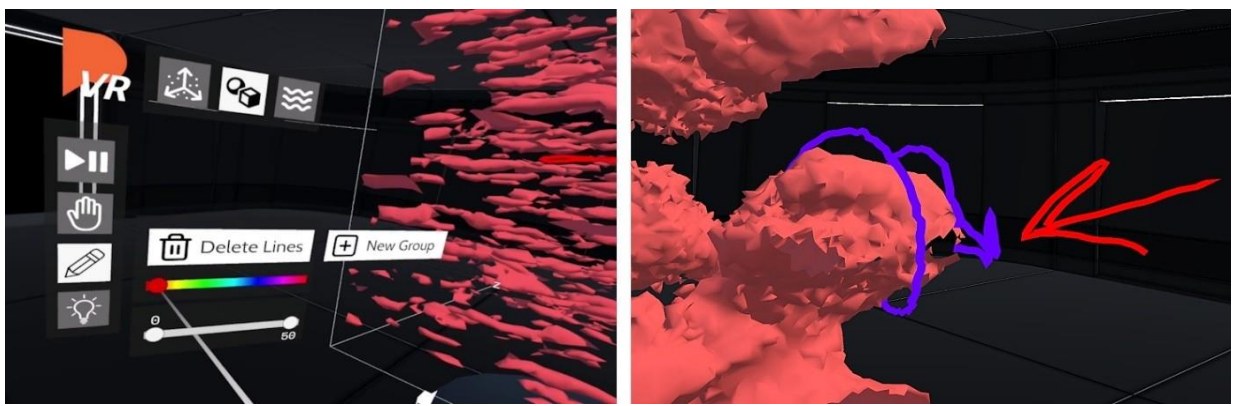

**Figure 9** The annotations menu allows users to define the annotation's characteristics, including the range of animation frames where they will be shown (left). Users can do 3D annotations to highlight relevant portions of the idioms (right).

This feature is also adapted to the dynamic nature of the plasma idiom. Because the plasma structure representation changes with time, the users can select a specific time frame where the annotation will be visible. This allows researchers to make dynamic annotations that track, for example, the path of a particular anomaly as the plasma animation progresses (Figure 10).

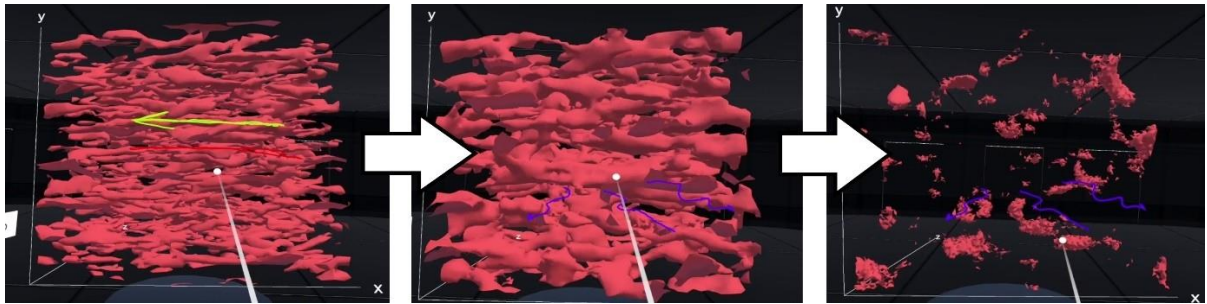

**Figure 10  Multiple annotations for different timeframes within the same simulated field.**

During the first iterations of the development of *PlasmaVR,* the analysis was carried out in an empty environment, allowing users to analyze the idioms without distractions. Although this helped highlight the data, the VR experience suffered due to the environment's lack of spatial reference points. As such, the surrounding environment was modeled as a sci-fi laboratory. However, to retain the empty background's advantages, a 'dark mode' was implemented that allows the user to switch off the lights of the surrounding environment while keeping the idioms illuminated.

# 4 Evaluation

To support the development of *PlasmaVR*, a preliminary version of the prototype was demonstrated to a group of five domain experts from GoLP. This initial step aimed to obtain feedback concerning the general perception of the application's interface and the immersive environment characteristics. It consisted of showing pre-recorded videos of the interaction with a plasma field dataset of medium size (2883 field size). This interaction included manipulation and navigation aspects such as rotation and locomotion inside the virtual environment. At the end of the presentation, a discussion with the participants was held about the prototype's characteristics and objectives. From this assessment, some key insights were collected. Among others, researchers reacted positively to the visualization possibilities enabled by *PlasmaVR*, stating that the tool aligned with the look and feel they desired from a plasma field analysis application. Based on the data collected, the prototype's functionalities were further revised.

To evaluate the revised version of the prototype, domain experts from GoLP interacted with the application and performed a set of predefined tasks. An array of objective metrics was recorded during this interaction. They were then asked to answer usability questionnaires concerning the prototype's usability. This section details how that evaluation was carried out.

## 4.1 Methodology

The study was conducted with an experimental group of five domain experts from GoLP, from which informed consent was obtained. As researchers from this research department, the participants were ideal to measure the possible benefits of *PlasmaVR* in the visualization of the data resulting from plasma physics simulations. The study occurred in a reserved room at the GoLP facilities at IST. The space was quiet, and because it was integrated into their workplace, it allowed the participants more convenient access to the testing sessions. In each session, only the participant and the observers were in the room. Initially, the participants were welcomed to the testing setup and given a consent form. They were then offered some introductory context to the objectives of *PlasmaVR* and how the testing session would take place. That introduction used an instructional video to ensure every participant received the same information. They then completed a participant characterization questionnaire with demographic data, previous VR technologies experience, and their background in plasma physics simulations.

After completing these initial steps, the participants were ready to interact with *PlasmaVR*. Before proceeding to the actual test tasks, they were allowed to interact freely with the prototype for a few minutes. This initial interaction allowed them to explore and understand how the controls worked and what features they could use. Then, when they felt ready, they were asked to perform tasks that simulated real-life usage scenarios. The hardware consisted of an *Oculus Quest 2* VR headset with a pair of controllers. The headset was connected to a desktop computer with an *Intel Core i7-8700* CPU @ 3.20GHz processor, 16GB of RAM, and a *NVIDIA GeForce GTX 1060* 3GB graphics card. A monitor, keyboard, and mouse were also used (for filling out questionnaires).

The participants performed the tasks in a sitting position (Figure 11), limiting the area they could reach to avoid accidents. To hinder the inconvenience and loss of immersiveness associated with removing the VR headset between tasks, the descriptions of the tasks were sequentially included in a panel inside the VR environment. The participants were asked to perform a set of three tasks that were designed to be realistic and, at the same time, would allow them to have contact with the extended scope of the prototype's functionalities. In the first task, the participants were required to draw a wide circle of a specific color surrounding a pre-drawn isosurface idiom and then define a time interval for when the annotation would be visible in the simulation. This task was aimed directly at testing the performance of the annotation and the corresponding timeline. Namely, it was introduced to assess the spatial perception within the immersive environment by testing how easy and intuitive it was to use the annotation tool to draw in space. It equally allowed the assessment of how the participants would perform in changing the color of an annotation element and specifying a timeline interval within the simulation using the interface panels.

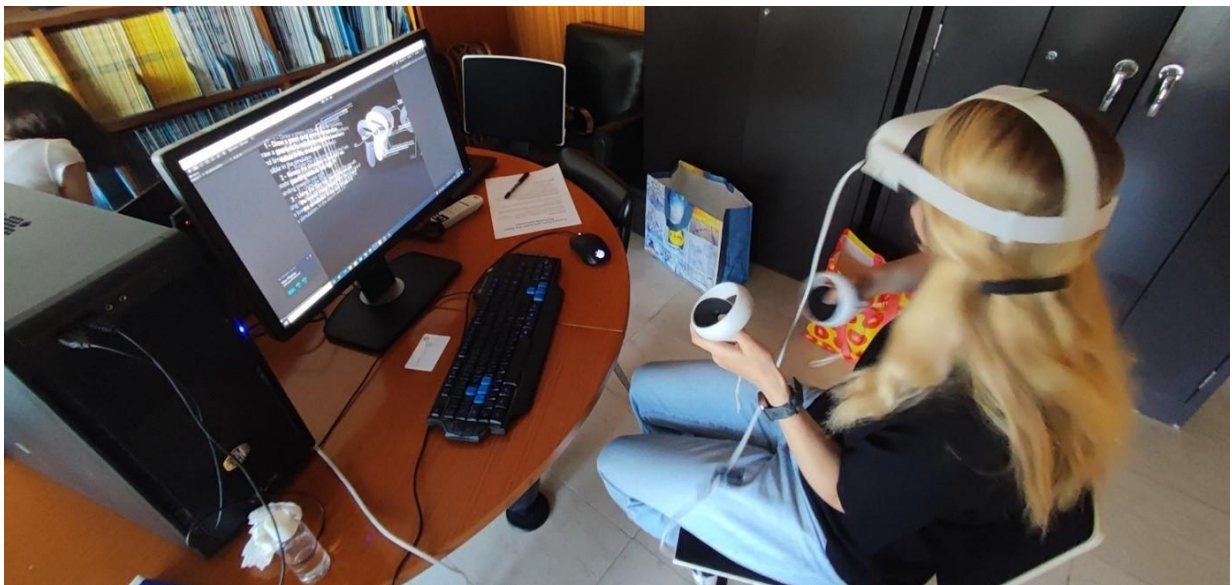

**Figure 11 User testing** *PlasmaVR***.**

For the second task, the participants were asked to rotate the isosurface to have the y-axis pointing towards the right. This more straightforward task was aimed at testing the rotation functionality's performance. In particular, in assessing how the proxy rotation would perform in terms of precision in relation to the VR controller's movement. This performance would indicate how well depth perception was transmitted to the user while using the rotation mechanism.

In the third task, the participants were asked to find the portion of an isosurface with the lowest charge density value. This task was designed to assess the analytics performance of the prototype. It required participants to use the slicing tool to extract the energy heatmaps to be able to figure out the portion with the lower density. Such a sequence of operations provided an adequate estimation of how well spatial relationship was perceived in the isosurface visualization. Likewise, it provided a good indicator of how well users can match this 3D visualization of data with the two-dimensional representation in slices.

After completing the three tasks, the participants removed the VR headset and were asked to fill out two questionnaires. The first was a 10-item System Usability Survey (SUS) [39] questionnaire. The second was an additional feedback questionnaire for the participants to leave comments, suggestions, and rate specific aspects of VR interaction (e.g., motion sickness, discomfort).

## 4.2 Results and discussion

The group of five participants was composed of researchers with ages ranging from 18-50 years old. All had experience with plasma simulations and OSIRIS/PiC Codes (20% more than five years, 60% between one and five years, and 20% less than one year experience). Most (80%) had earlier experience with using VR technologies.

To assess the suitability of the prototype for different types of tasks, a set of objective metrics, such as completion time and number of errors, were analyzed. The time participants took to complete each task type (annotation, rotation, and slicing, as described in the previous section) is shown in Figure 12. Both the annotation and the slicing tasks required a greater number of interactions from the participant to achieve the proposed goals than the more straightforward rotation task. As such, and as expected, the rotation task corresponded to a shorter (mean time ± standard error) completion time ($1.32 \pm .153$) when compared to annotation ($2.87 \pm .321$) and slicing ($3.50 \pm .224$).

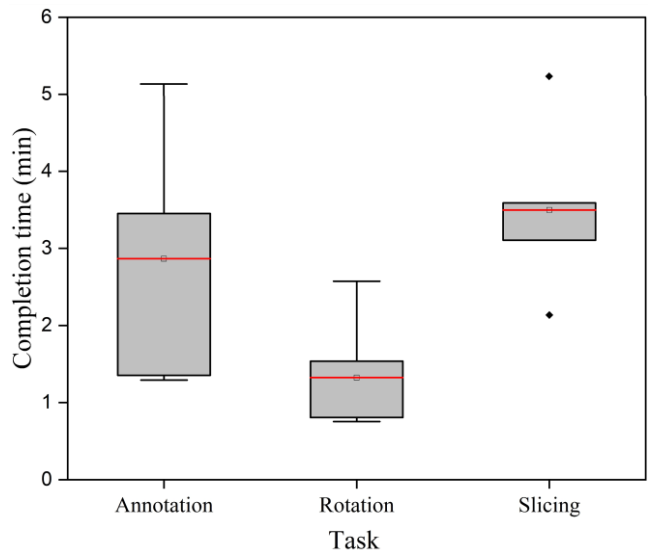

**Figure 12** Comparison between the time needed to complete each task type.

If we also take into consideration the number of errors made by participants in each task (shown in Figure 13), we can see that while the slicing task had a marginally higher completion time, it corresponded to a much lower (mean number of errors ± standard error) number of mistakes (0.60 ± .179) than the annotation task (1.80 ± .219). This difference may indicate that the prototype is better fitted to handle analytics tasks than annotation tasks. The rotation task had a similar number of errors (0.60 ± .110) to the slicing task.

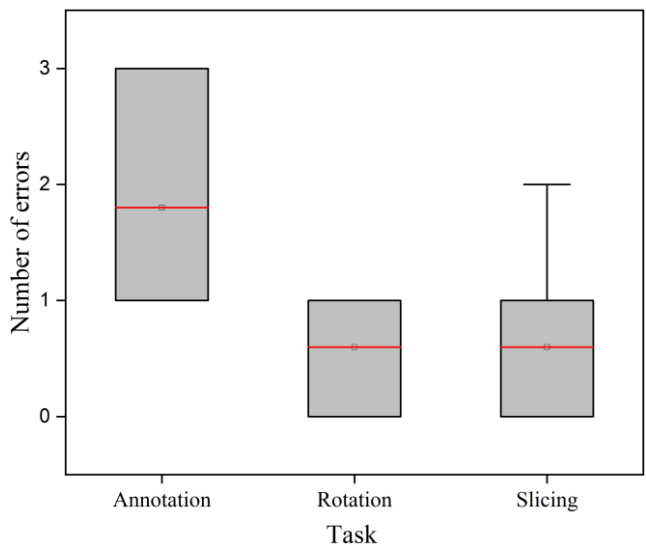

**Figure 13** Comparison between the number of errors made by participants for each task type.

The participants were able to complete the three tasks with different levels of success. This means that sometimes, while the end goal was achieved, not all the task requirements were fulfilled (e.g., an annotation was effectively made but in the wrong place or with an incorrect color). So, we also wanted to understand how the different steps and sub-goals for each task were achieved so that the aspects with which participants had the most difficulties could be more accurately addressed in future developments. In Figure 14, we can observe the completion of each individual step for the three tasks. For the annotation task, most of the uncompleted steps are related to selecting the time interval for when the annotation would be visible

(40% had difficulties finding the timeline tool, and 60% were unable to set the time frame correctly). A significant percentage of users (80%) also failed to draw the circle in the correct position. Some of the errors registered in this task resulted from the users being confused about where to find the annotation feature in the interface. So, how this feature was integrated into the interface is an aspect to consider for future prototype improvement. Regarding the rotation task, while all participants reached the end goal, some (60%) had difficulties correctly understanding the interaction mechanism associated with the idiom rotation. Lastly, for the slicing task, uncompleted steps resulted mainly from participants having problems obtaining the range of slices for a specific time step (40%) and identifying and accurately selecting individual slices (20%).

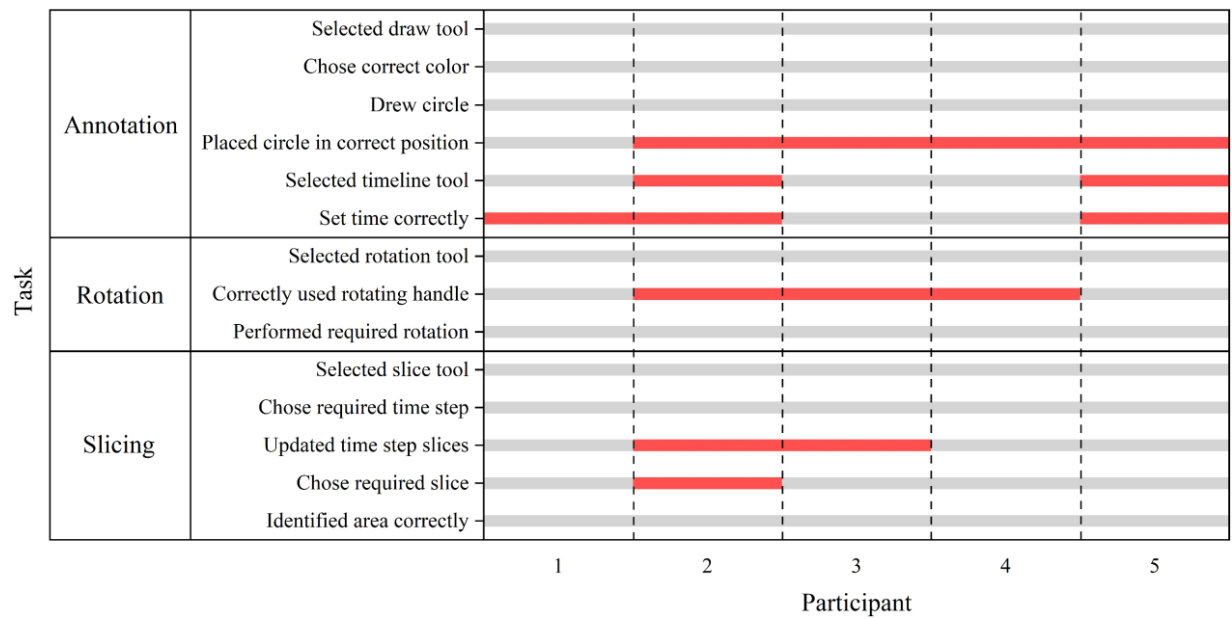

**Figure 14** Completion distribution for the different task steps. The steps that participants did not successfully complete are highlighted in red.

To assess the usability of *PlasmaVR* (**RQ1**), the participants were asked to complete a ten-question standard SUS questionnaire with a five-level *Likert* scale for agreement (1: Strongly disagree and 5: Strongly agree) after interacting with the prototype. An average SUS score of 75.5 (SD = 5.5) was obtained. Such a score can be paired with a rating of 'Good'[40] or 'B' (74.1 - 77.1)[41]. Although the small number of participants impacts the accuracy of the usability evaluation (< 35% accuracy, based on the sample size threshold proposed by Tullis and Stetson[42]), the majority had a positive perception and found the system usable.

We also wanted to identify the specific aspects where the system performed well and where it could be enhanced in the scope of future development efforts. With that objective, the SUS questions were broken down into eight categories. These reflect the different usability areas that were addressed. The categories include cohesiveness (how well-integrated the prototype's features are), learnability (how easily it can be learned), and intuitiveness (how simple and easy to use). They also include concision (how uncomplicated the interface is), reliability (how few inconsistencies are in the prototype), and comfort (how non-frustrating its use is). Furthermore, they include trustworthiness (how confident using the prototype participants were) and usage intention (how much participants expected to use it). Results show that the aspects that performed better (mean SUS score ± standard error) were cohesiveness (3.40 ± .110), learnability (3.33 ± .067), intuitiveness (3.20 ± .167), concision (3.20 ± .167), and reliability (3.20 ± .089). Aspects that performed below average include comfort (2.80 ± .089) and trustworthiness (2.60 ± .110). The lowest contribution came from usage intention (1.80 ± .089). These results are illustrated in Figure 15.

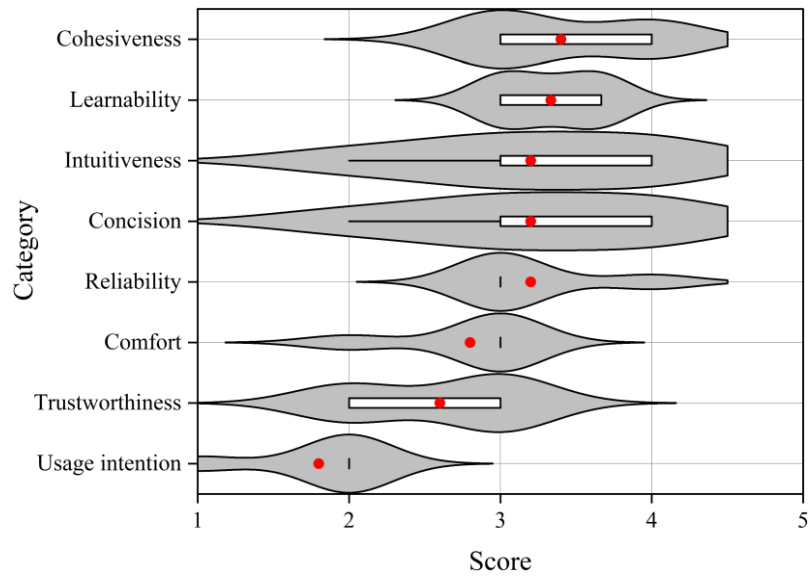

**Figure 15** Distribution of participant's score for each usability category.

The results are mainly within a relatively narrow range (mean scores between 2.5 and 3.5). The aspects that performed better are those related to the overall experience with the prototype and how simple and easy to use the system is. Participants found the interface reliable, well-structured, and cohesive. A good level of learnability was also observed, which seems to be a direct consequence of how intuitive the participants considered the interface. These results align with the participants' increased perception of the usefulness of VR after testing the prototype, which will be addressed further on. Nevertheless, participants found that while the prototype was reliable, they did not feel entirely confident using it. This apparent contradiction might be justifiable by interaction limitations pointed out by participants (e.g., difficulties in doing a full rotation of the idiom using a single controller motion, which will be discussed later). The most unexpected result, however, was the one corresponding to usage intention, which achieved the lowest score of all categories. This lower classification might result from the more localized role in their workflow that some researchers identified as the main scope of the prototype. More specifically, some participants positioned its usefulness in a faster preliminary visualization of the plasma physics data as a first step before moving to a more exhaustive analysis using conventional means.

Additionally, we wanted to find if the participant's perception of the usefulness of VR in plasma simulations (**RQ2**) had changed after experiencing the prototype. Before using the prototype, the participants were divided on the usefulness of VR in plasma simulations. Indeed, only 40% agreed that VR would be useful in helping their workflow, 20% neither agreed nor disagreed, and 40% disagreed (five-level *Likert* scale for agreement, 1: Strongly disagree and 5: Strongly agree). After using the prototype, participants' perception changed to much more positive values, with 60% strongly agreeing on the usefulness of VR in plasma simulations, 20% agreeing, and 20% neither agreeing nor disagreeing (Figure 16). This substantial shift in opinion regarding the usefulness of VR in plasma simulations suggests that the user experience with the prototype was impactful and meaningful to the participants. As such, it is consistent with a high level of engagement despite the lower usage intention previously addressed.

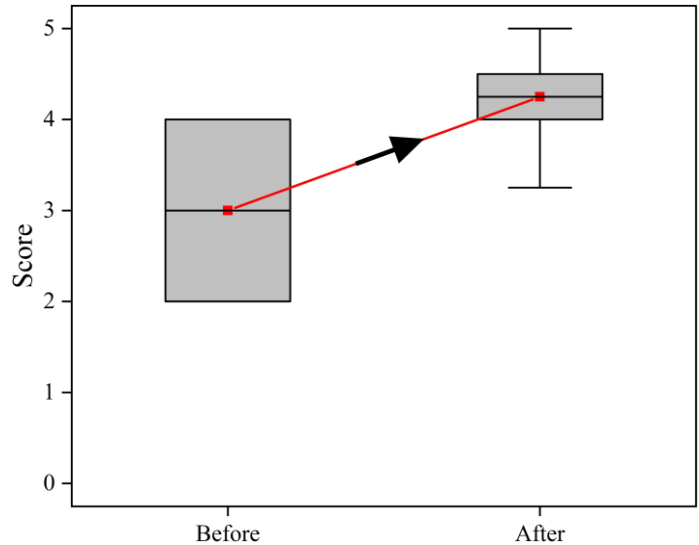

**Figure 16 Participants' perception of the usefulness of VR in plasma simulations before and after experiencing the prototype.**

We also enquired users about the aspects of the prototype they thought would contribute to making VR application useful in their workflow (Figure 17). 20% strongly agreed, and 80% agreed that VR could be used to display data effectively. 60% strongly agreed, 20% agreed, and 20% neither agreed nor disagreed that VR was useful for highlighting relevant information. 20% strongly agreed, 40% agreed, 20% neither agreed nor disagreed, and 20% disagreed that VR improved the data exploration experience. Lastly, 80% strongly agreed, and 20% agreed that VR was a more enjoyable way of exploring plasma simulation data than conventional means. It's worth highlighting that the highest of these scores corresponds to the way participants found *PlasmaVR* enjoyable, which is, once again, compatible with a high level of engagement.

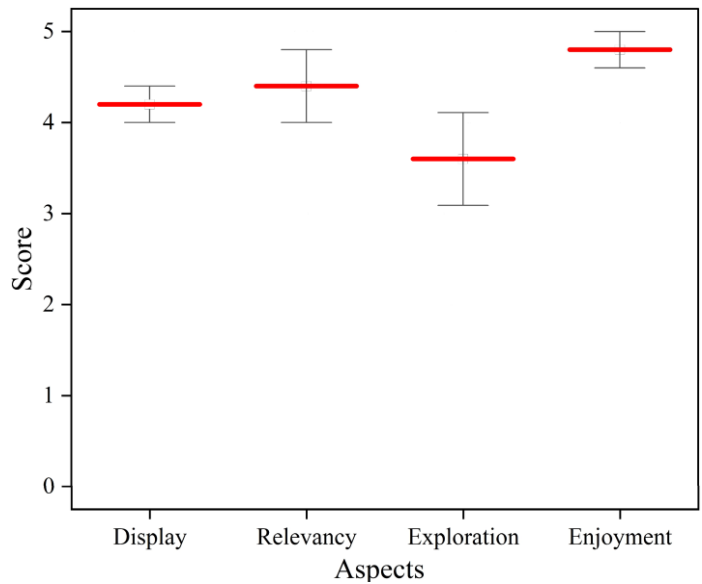

**Figure 17 Aspects of the prototype that participants considered would contribute more to making the use of VR useful in their workflow.**

The additional feedback questionnaire also allowed participants to complement their feedback with comments and suggestions. Regarding the enjoyability of the VR experience, participants left comments such as 'the immersion is really fun' and that what they liked the most was 'the interactive and fun way of using it'. Regarding the data exploration experience, some commented that they could 'quickly check some data information (like slices), which sometimes takes time [using a computer]' and highlighted the 'easy control of what [they wanted] to see in the simulation'. Concerning the effectiveness of data display, a participant stressed the 'better intuition of what's going on in the simulation' in the VR experience. About the ability to present relevant information, participants commented that they 'really enjoyed [seeing] the data well represented in 3D'. Some participants underlined that the prototype can also be useful to promote engagement in scientific dissemination activities. Others emphasized the usefulness of the prototype as a tool for a preliminary visualization of the data before moving to a more exhaustive analysis. In that scope, they highlighted the benefits of VR's increased spatial relationship perception over conventional visualization means, like 2D screens.

Conversely, the participants were also asked to comment on what they found more frustrating about the VR experience. Some participants stated that sometimes they were 'unable to see immediately what the tools [were] for'. Others wrote that the animations were 'playing very fast', making it 'hard to follow what [was] happening', and suggested that 'having more frames would help'. A participant reported that the movement through the environment felt slow, and it would be helpful to have the possibility of sprinting through the environment. Regarding object manipulation, a participant noted that turning the whole idiom using a single motion was difficult. Lastly, a participant highlighted that 'adding more post-processing functions of the fields' would greatly benefit the prototype analysis capacities.

With the hardware used, the prototype maintained a satisfying and consistent graphical performance of around 90 average frames per second (fps) throughout the several stages of the simulation. The most demanding phase (the generation of the slice planes) corresponded to a drop of just a couple of fps below the average value. This consistency could have contributed to the positive user experience. Indeed, when the participants were asked if they had felt any physical discomfort using the application (including motion sickness or head discomfort), 80% answered that they had felt no discomfort at all, and 20% wrote down that they had felt some discomfort.

### 4.3 Limitations and future work

When analyzing the results from this work, the rationale behind a few methodological decisions and their corresponding limitations should be considered. The first of these limitations is the small sample size used in the study, which makes it harder to assert statistical significance. In that sense, the study could have been carried out with more participants, namely by extending the sample selection to encompass non-expert users or physics researchers from areas other than plasmas. We chose instead to restrict the sample selection to domain experts, which allowed us to ensure a consistent baseline among the participants regarding plasma experiments data analysis knowledge. Likewise, we opted for using a more controlled experimental environment instead of, e.g., making the application available online and asking users to download and execute the application by themselves.

Another methodological issue that limits the scope of this study is the absence of a comparative assessment of conventional analysis means with VR, namely by using an experimental control group. As such, assumptions regarding improvements in performance or usability can only be substantiated by participants' feedback (collected using questionnaires) and not by differences in measured performance metrics. Nevertheless, while not in the scope of this study, such comparative assessment can be addressed in future work.

There are, in fact, several opportunities to expand this study beyond its current scope. Potential future research directions may include overcoming some of the limitations mentioned above. Such improvements may mean extending the experimental group to include researchers from plasma physics research units other than GoLP. They may also imply, e.g., the comparison of *PlasmaVR* with augmented reality and desktop versions of the prototype to further assess the advantages and limitations of immersive environments in plasma physics simulations.

Future research efforts may also focus on the potential for *PlasmaVR* to be applied in scientific outreach. As noted by some participants, specific improvements can be made to adjust the prototype to such a task better. One of those suggested improvements was the inclusion of pre-rendered animations in the environment, showing simulation steps that cannot otherwise be displayed in real time. Other suggestions in this scope included adding audio and text to contextualize the simulations for non-experts further.

## 5 Conclusion

This work presents a novel prototype tool for visualizing plasma physics experiment datasets in VR. The tool enables a multidimensional data visualization environment where users can travel around, and inside animated representations of plasma based on time-dependent data. It allows them to observe the structural variations in morphology over time from several points of view. The work shows the different characteristics of the tool, including its architecture, raw data processing capabilities, and user interface functionalities. It addresses its evaluation with a group of domain experts consisting of plasma physics researchers. This evaluation is carried out using a set of objective and subjective metrics. These metrics are collected during testing through direct measurement, and after testing using questionnaires. The collected metrics are used to support the research questions, which aim to ascertain the usability and usefulness of VR in plasma physics visualization. The findings suggest that applying VR technologies to plasma physics visualization can result in a usable experience. The results also support the hypothesis that VR can be useful in plasma physics visualization. In future, collaboration between multiple participants by contemplating collocated proximal interactions, proxemics [44], would be an interesting feature to contemplate.